\documentstyle[aps,prl,preprint]{revtex}
\begin{document}
\tighten
\title{Anomalous density of states of a Luttinger liquid in contact with
	a superconductor}
\author{  C. Winkelholz$^{(1)}$, 
	  Rosario Fazio$^{(2)}$,
	  F.W.J. Hekking$^{(3)}$, 
	  and 
	  Gerd Sch\"on$^{(1,4)}$}
\address{
 $^{(1)}$Institut f\"ur Theoretische Festk\"orperphysik,
	 Universtit\"at Karlsruhe, 76128 Karlsruhe, Germany\\
 $^{(2)}$Istituto di Fisica, Universit\`a di Catania,
	 viale A. Doria 6, 95129 Catania, Italy\\
 $^{(3)}$Cavendish Laboratory, University of Cambridge, Madingley Road,
         Cambridge CB3 0HE, UK\\
 $^{(4)}$Department of Technical Physics, Helsinki University of
	Technology, 02150 Espoo, Finland\\}
\maketitle

\begin{abstract}
We study the frequency and space dependence of the 
local tunneling density of states of a Luttinger liquid (LL) which
is connected to a superconductor. This coupling {\em strongly} modifies
the single-particle properties of the LL.
It significantly enhances the density of states near the 
Fermi level, whereas this quantity vanishes as a power law for 
an isolated LL.
The enhancement is due to the interplay between electron-electron
interactions and multiple back-scattering processes of low-energy
electrons at the interface between the LL and the superconductor. 
This anomalous behavior extends  over large distances 
from the interface and may be detected by coupling normal probes to
the system.

\end{abstract} 


\narrowtext

Transport in low-dimensional structures is strongly influenced by
electron-electron interactions. A paradigm model to describe interactions in
one-dimensional systems is the Luttinger liquid (LL). The low-lying
excitations of the electron system consist of charge and spin  waves, rather
than quasiparticles~\cite{Emery,Goni}. As a consequence,  the
presence of a barrier in the liquid leads
to perfectly reflecting (for repulsive  interactions) or   transmitting (for
attractive interactions) behavior at low  energies~\cite{Kaneetal}.  

One of the most striking characteristic properties of Luttinger liquids
is the  behavior of the density of states (DOS) close to the Fermi energy.  
Contrary to Fermi liquids, whose  quasiparticle residue is finite, 
LLs have a DOS which vanishes near the Fermi energy as a power law,
\begin{equation}
	N(\omega ) \sim  \omega ^{(g_{\rho}+4/g_{\rho}-4)/8} .
\label{LLDOS}
\end{equation}
The exponent $g_{\rho}$ depends on the strength of the
electron-electron interaction: it is smaller (larger) than two  for 
repulsive (attractive) interactions.  In the non-interacting case 
$g_{\rho} = 2$, and the DOS is constant
as in Fermi liquids~\cite{Voit}.

Recently it became possible to fabricate interfaces between a
superconductor (S) and a two-dimensional electron gas (2DEG)~\cite{Karlsruhe}. 
An excess low-voltage conductance due to Andreev scattering  
has been observed in Nb-InGaAs 
junctions~\cite{Kastalsky}, as well as a supercurrent 
through a 2DEG in an InGaAs/InAlGaAs heterostructure with Nb
contacts~\cite{Nitta,Dimoulas}. If the 2DEG is gated to form a quantum 
wire, it should be possible to study transport through 
Superconductor - Luttinger Liquid  (S-LL) interfaces. 
The Josephson current through a S-LL-S 
system has been calculated~\cite{Fazio,Maslov}, as well as  the $I$-$V$
characteristics of a tunnel junction between a superconductor and a chiral
LL~\cite{Fisher}. The latter can be  realized, e.g., in the fractional 
quantum Hall regime. 

The proximity effect~\cite{Deutscher} modifies the properties of a 
normal metal (N) in contact with a superconductor.
The leakage of Cooper pairs induces a non-vanishing  pair amplitude in N, 
defined as 
$
F(\vec{r})=\langle \psi _{\uparrow}(\vec{r})
\psi _{\downarrow}(\vec{r}) \rangle
$,  
where $\psi _s(\vec{r})$ is the annihilation operator for an electron with 
spin $s$.
The pair amplitude is  a two-particle property, related to the probability 
to find two time-reversed electrons at a position $\vec{r}$. 
In a clean normal metal at zero temperature, $F(\vec{r})$ decays as $ 1/r$
away from the N-S interface. In a LL with repulsive interaction
in contact with a superconductor,
$F(\vec{r})$ decays as $1/r^{\gamma}$, where
$\gamma > 1$ depends on the strength of interaction~\cite{Maslov}. 
These results hold as long as $\vec{r}$ lies within the temperature-dependent
coherence length $\xi _N = \hbar v_F /k_BT$ from the interface 
to the superconductor. At larger distances,
$F(\vec{r})$ decays exponentially on the length scale 
$\xi _N$. The reason for the decay of the pair amplitude is that the
two electrons
loose their {\em relative} phase coherence over this distance. {\em Single}
electrons, however, loose phase coherence only at  a much larger distance,
namely the phase-breaking length $L_{\phi}$. Indeed, recent
experiments~\cite{Dimoulas,Courtois} have shown that interference  effects 
due to single quasiparticles in N-S systems persist over distances much 
larger than $\xi _N$. We, therefore, expect quite generally a 
considerable influence of superconductivity on  {\em single particle}
properties over distances where the pair amplitude has already decayed.
 
In order to investigate these properties in a strongly interacting system
we study the local single-particle density of states 
(DOS) of a LL in contact with a superconductor. We find that the local DOS is 
substantially {\em enhanced} near the Fermi energy as compared to the power 
law decay of an isolated LL (cf. Eq.(\ref{LLDOS})). This result
should be contrasted with the behavior of the pair amplitude (a two-particle
property) which is {\em suppressed} in the interacting system.
The anomalous enhancement is a result of the interplay between  the 
scattering of low-energy electrons at the S-LL interface and the  
electron-electron  interactions in the LL~\cite{Oreg}. 
As for the space dependence, the DOS does not  decay in the same
fashion as the pair amplitude away from the S-LL interface.  
It remains enhanced up to distances of the order of the mean
free path, which may be much larger than $\xi _N$ for a clean quantum wire. 
Hence, the effect cannot simply be explained in
terms of a finite  density of Cooper pairs in the  LL. 
Both the frequency and the space dependence of the DOS can be detected 
experimentally by coupling normal metal tunneling probes to the LL at some
distance from the superconductor.

In the inset of Fig.~\ref{setup} we schematically draw a 'clean' S-LL 
interface, consisting of a LL in good contact with S.
The shaded area indicates a tunnel junction between the LL and a normal 
metal probe. We will also study a LL of  finite length connected to two 
superconductors (S-LL-S system). In such a system  Andreev bound states 
exist below the gap~\cite{Kulik}. Finally, we will comment 
on the case of a LL connected to S by means of a tunnel barrier.  
The anomalous enhancement of the DOS is found in this case as well.
  
The Hamiltonian of a LL can be written in bosonized form as ($\hbar =1)$
\begin{equation}
	\hat{H}_L = \frac{1}{2}  \sum _{j} v_j \int dx 
	\left[ \frac{g_j}{2} (\nabla \phi _j )^2 +
	\frac{2}{g_j} (\nabla \theta _j)^2
	\right]\; ,
\label{lutham}
\end{equation}
where  $j=\rho,\sigma$, and  $v_{j}=(2/g_{j})v_{F}$ are the
renormalized interaction-dependent Fermi velocities. 
We restrict ourselves to repulsive, spin-independent
interactions; hence $g_{\rho}<2$ and $g_{\sigma}=2$. 
The Fermi field operators are
decomposed in terms of right- and left-moving Fermion operators 
$\psi_{+,s}$ and $\psi_{-,s}$, respectively, 
$
\psi_{s} =e^{ ik_{F}x}\psi_{+,s} + e^{- ik_{F}x}\psi_{-,s} 
$,
where $k_F$ is the Fermi wave vector. 
The fields $\psi_{\pm,s} $ in turn can be expressed through Boson operators
\begin{equation}
	\psi_{\pm,s}^{\dagger}=\sqrt{\rho_{0}}
	e^{i\sqrt\pi
	[\pm \phi_{s}(x)+\theta_{s}(x)]} ,
\label{boseform}
\end{equation}
where $\theta_{s}=\frac{1}{\sqrt{2}}(\theta_{\rho}+s\theta_{\sigma})$ and 
$\phi_{s}=\frac{1}{\sqrt{2}}(\phi_{\rho}+s\phi_{\sigma})$. The density of
electrons per spin in the LL is $\rho_{0}=k_{F}/2\pi$. 
Maslov et {\it al.}~\cite{Maslov} recently developed a  bosonization 
scheme to treat  clean S-LL interfaces. For a LL coupled  to two 
superconductors at a distance  $L$, they  obtained the 
following normal mode expansion for the the fields 
\begin{eqnarray}
	\theta _{\rho}(x)	
	& = &
	\sqrt{\frac{\pi}{2}} (J+\chi)\frac{x}{2L} +
	\frac{i}{2}  \sqrt{\frac{g_{\rho}}{2}} \sum _{q > 0}
	\gamma_q
	\sin(qx)
	(\hat{b}^{\dagger}_{\rho,q}
	-
	\hat{b}_{\rho,q}) ;\label{phitheta1}\\
	\theta _{\sigma}(x)	
	& = &
	\frac{1}{\sqrt{\pi}}\theta^{(0)}_{\sigma} +
	\frac{i}{2} \sqrt{\frac{g_{\sigma}}{2}} \sum _{q > 0}
	\gamma_q
	\sin(qx)
	(\hat{b}^{\dagger}_{\sigma,q}
	+
	\hat{b}_{\sigma,q}) ;\label{phitheta2}\\
	\phi _{\sigma}(x)	
	& = &
	\sqrt{\frac{\pi}{2}} M\frac{x}{2L} +
	\frac{i}{2}  \sqrt{\frac{2}{g_{\sigma}}} \sum _{q > 0}
	\gamma_q
	\sin(qx)
	(\hat{b}^{\dagger}_{\sigma,q}
	-
	\hat{b}_{\sigma,q}) ;\label{phitheta3}\\
	\phi _{\rho}(x)	
	& = &
	\frac{1}{\sqrt{\pi}}\phi^{(0)}_{\rho} +
	\frac{i}{2} \sqrt{\frac{2}{g_{\rho}}} \sum _{q > 0}
	\gamma_q
	\sin(qx)
	(\hat{b}^{\dagger}_{\rho,q}
	+
	\hat{b}_{\rho,q}) \:\:.
\label{phitheta4}
\end{eqnarray}
Here, $\hat{b}^{(\dagger)}_{j,q}$ are Bose operators and
$\gamma_q = \exp\{-q\alpha/2\pi\}$ where $\alpha$ is a short range cut-off. The
phase difference between the two superconductors is $\chi $; $J$ and $M$ 
describe the
topological excitations satisfying the constraint $J+M=$ odd.  Finally,
$\theta^{(0)}_{\sigma } $ and $\phi_{\rho}^{(0)}$ are canonically conjugate to $M,J$.
The local density of states (per spin) of the LL measured at a distance $x$ from
the superconducting contact is obtained from the retarded
one-electron Green's function of the LL,
$G_R (x,x';t) \equiv -i\langle \{\psi_s(x,t),\psi_s^{\dagger}(x',0)\}\rangle
\theta (t)$,
\begin{equation}  
	N(x,\omega ) = - \frac{1}{\pi} {\cal I} \mbox{m} 
	\int _{-\infty} ^{\infty} dt 
	e^{i\omega t} G_R(x,x;t) \:\:\:\:\:\: .
\label{DOS}
\end{equation}

We first discuss the space and frequency dependence of the DOS of a LL
contacted at $x=0$ with a superconductor, which corresponds to
the limit $L \to \infty$ in the mode expansion given by 
Eqs.~(\ref{phitheta1}) -- (\ref{phitheta4}).  
In this case only the non-zero modes ($q>0$)
contribute to the local DOS. The correlation function
$\langle \psi_s^{\dagger}(x,t)\psi_s(x,0)\rangle$ can be evaluated
using the boson representation Eq.~(\ref{boseform}) with the result
\begin{eqnarray}
\langle \psi_{s}^{\dagger}(x,0) \psi_{s}(x,t) \rangle 
	 = 
        2\rho_0\prod_{j=\rho,\sigma} & 
	\left(\frac{\alpha ^2 + (2x)^2}{\alpha ^2}\right)^{\gamma_{j}}
	\left(\frac{\alpha ^2}{(\alpha-iv_{j}t)^2}
	\right)^{\eta_{j}} \nonumber \\
	& \times  \left[\frac{\alpha ^2}{(\alpha -i(2x+v_{j}t))
	(\alpha +i(2x-v_{j}t))}
	\right]^{\gamma_{j}} \;,
\label{correlator}
\end{eqnarray}
at a distance $x$ from the LL-S interface, 
where $\gamma_{j}=(g_{j}/16-1/(4g_{j}))$ and $\eta_{j}=(g_{j}/16+1/(4g_{j}))$.
At small energies the DOS behaves as
\begin{equation}
	N_{S-LL}(\omega ) \sim \omega^{g_{\rho}/4-1/2} .
\label{LLSDOS}
\end{equation}
The exponent of the DOS is negative ($g_{\rho} < 2$), which implies
a {\em strong  enhancement} at low energies whereas in the absence of S 
the DOS of the  LL {\em vanishes} at the Fermi energy. 
The presence of the superconductor thus changes 
the  properties of the Luttinger liquid in a qualitative way. Recently, Oreg
and Finkel'stein~\cite{Oreg} have found a similar enhancement 
of the local DOS of a 
LL in the presence of an impurity. They interpret their
result as a consequence of the interplay between the back-scattering 
induced by the impurity and the repulsive interactions in the LL. 
A similar interplay exists in our system. 
Although we consider a clean S-LL interface,
backscattering is induced by the superconducting gap, 
which reflects low-energy 
electrons either directly or via (multiple) Andreev processes.
The enhanced DOS as a function of frequency, Eq.~(\ref{LLSDOS}),
is schematically drawn in Fig.~\ref{setup}; for comparison we also show 
the vanishing DOS in absence of the superconductor, Eq.~(\ref{LLDOS}). 

At low energies $\omega$ the enhancement of the DOS persists over large 
distances $x(\omega) \sim v_{\rho}/\omega $ from the interface. 
On the other hand, the induced pair amplitude in the LL, 
which is characteristic of the presence of the superconductor, 
decays as a  power~\cite{Maslov} of the distance $x$.  
This profound difference in the space dependence demonstrates that the DOS
provides different information compared to the  proximity effect. 
The reason why the DOS does not approach the well-known behaviour of an
Luttinger liquid far from the superconducting contact is in part 
related to the fact that we are considering a clean wire. In this case the 
states in the LL are extended and the DOS enhancement does not depend on $x$. 

We now turn to the properties of the DOS for a S-LL-S system. The two 
superconductors are separated by the distance $L$ and are kept at a phase 
difference $\chi$. The latter can be achieved, e.g., by embedding
this junction in a SQUID. As we consider a LL of finite length,
the topological excitations should be taken into account; moreover the  
contribution from the non-zero modes consists of a discrete sum rather than
a continuous integral over $q-$states. 
The correlator reads
\begin{equation}
\langle \psi_{\pm,s}^{\dagger}(x,0) \psi_{\pm,s}(x,t) \rangle 
	= 	
	e^{i\pi v_F(1\pm\chi/\pi)t/2L} \prod_{j=\rho,\sigma}D_{j}(x,t)  . 	
\label{correlator1}
\end{equation}
The exponential prefactor  originates from  the topological part; we used the
fact that $J=1$ and $M=0$ in the ground state for $-\pi < \chi < \pi$. The
$\chi$-dependence is related to the phase-dependent shift of the Andreev
levels (see below). The contribution from the non-zero  modes is given by

\begin{eqnarray}
	D_{j}(x,t)
	 =  	
	& \rho_{0} &
	\left(\frac{(1-e^{-\pi(\alpha+2ix)/L})(1-e^{-\pi(\alpha-2ix)/L})}
	{(1-e^{-\pi\alpha/L})^2}\right) ^{\gamma_{j}} 
	\left(\frac{1-e^{-\pi(\alpha-iv_{j}t)/L}}
	{1-e^{-\pi\alpha/L}}\right)^{-2\eta_{j}} \nonumber \\ 
        & \times &
	\left(\frac{(1-e^{-\pi\alpha/L})^2}
                    {(1-e^{\pi[\alpha -i(2x+v_{j}t)]})
                     (1-e^{\pi[\alpha +i(2x-v_{j}t)]})}
	              \right)^{\gamma_{j}} \; .
\label{Dandreev}
\end{eqnarray}
As in the previous case of a single S-LL interface, the anomalous behavior of the DOS 
at low $\omega$ extends  over large distances,
(measured now relative to the position of one of the 
interfaces), hence 
the phase-dependent contribution to the DOS persists over distances much 
larger than the Josephson coupling. If the interaction constant can be 
written as a ratio of two integers ($g_{\rho}=m_0/n_0$), we can express the DOS, 
using Eq.~(\ref{correlator1}), as 
\begin{equation}  
	N(x,\omega ) = \rho_{0}\sum_{s,\pm}
            \sum_{n}a_{n}(x)[\delta(\omega-E_{\pm,n})
                           + \delta(\omega+E_{\pm,n})] \,.
\label{DOS1}
\end{equation}
Here $a_n (x)$ are the Fourier coefficients of the function $D(x,t)$
corresponding to the energies
$$
E_{\pm,n} =E_{F}(2\frac{n}{n_0}+1)\frac{\pi}{2Lk_{F}}\pm v_{F}\frac{\chi}{2L} \;\;\; ,
$$
where $E_F$ is the Fermi energy. The $E_{\pm,n}$ are the energies of the 
Andreev levels~\cite{Kulik} in the interacting quantum wire.
The phase-difference $\chi$ lifts the degeneracy for right- and left moving
electrons, giving rise to the Josephson effect. 
In the noninteracting case ($g_{\rho}=2$), all the $\delta$-functions have 
the same weight and the local DOS shows a peak whenever the frequency $\omega$ 
coincides with an Andreev level $E_{\pm,n}$.  
When the electron-electron interaction is switched on, the charge and 
spin part in  Eq.~(\ref{Dandreev}) obtain different periodicities due 
to  spin-charge separation. As a consequence the coefficients $a_n$ show a 
more structured behaviour.  In  Fig.~\ref{dosslls} the DOS is plotted for 
$g_{\rho} = 1$ as a function of the frequency and of the distance from one of 
the two  superconductors. For clarity we use a realistic, broadened 
version  of the $\delta$-functions in Eq.~(\ref{DOS1}).
We, further, 
 fixed the phase difference $\chi$ between the superconductors to zero.
One clearly sees a strong enhancement of the DOS close to the interface. 
Away from the superconductor the DOS remains enhanced, but the energy 
scale of the enhancement is reduced to lower frequencies. The oscillatory
contribution to the DOS is reminiscent of the Friedel-oscillations,
characterized by a period $2k_F$.
In the general case $\chi \ne 0$, the DOS for the right moving electrons differs from 
that of the left moving electrons due to the phase factor in Eq.~(\ref{correlator1}).
Although this leads to a more complicated dependence of DOS on $x$ and $\omega$, 
the anomalous enhancement is still present.  

So far we discussed the case in which the S-LL interface 
has a high transparency. Let us shortly comment on the opposite
limit, in which the Luttinger liquid is connected to the  superconductor by a tunnel
junction. In this case at low energies, 
we find for the DOS close to the junction
$N_{S-LL} \sim \omega ^{(g_{\rho}/2 -1) + (1/2g_{\rho} - g_{\rho}/8)}$. Although
the exponent is different from the one appearing in  Eq.~(\ref{LLSDOS}), the DOS
is clearly enhanced. Moreover, also in this case the enhancement is found regardless
of the distance from the junction.

In summary we considered the DOS of a Luttinger liquid  in contact with a
superconductor. We studied specifically the cases of  a single S-LL interface
and a S-LL-S system. Contrary to the well-known behavior in Luttinger liquids,
the presence of the superconducting contact  strongly {\em enhances} the local
DOS close to the Fermi energy, and this behavior extends to {\em large}
(energy-dependent) distances from the interface. 
Our results can be verified experimentally~\cite{Esteve}, 
e.g.,  by means of the setup drawn in the inset of Fig.~\ref{setup}. 
We imagine connecting the 
LL by means of  a tunnel junction to a normal metal (at a distance  $x$ 
from the interface)  and measuring the $I-V$ characteristic of this junction. 
If there were no superconductor  the conductance of the normal metal-LL  
junction would go to zero as  the temperature (voltage, frequency) is lowered.  
The presence of the superconductor leads to an excess conductance  at the junction.
     
\acknowledgments
We thank C. Bruder for numerous important suggestions 
and W. Belzig and  M. Fabrizio for useful discussions. 
The support of the Deutsche Forschungsgemeinschaft,
through SFB 195, the European Community through contract
ERB-CHBI-CT94-1764, and the A.v.Humboldt award of the Finish
Academy of Sciences (GS) is gratefully acknowledged.

\begin{figure}
\caption{Schematic dependence of DOS on frequency for a pure LL (dashed
	line) and for a LL connected to S (solid line).
	Inset: Luttinger liquid, connected adiabatically to a
	superconductor. The shaded area indicates a tunnel junction
	with a normal metal used to measure the DOS in the LL at a distance 
	$x$ from the interface.}
\label{setup}
\end{figure}

\begin{figure}
\caption{The local DOS for a S-LL-S system for particles of species 
	$p=\pm$, plotted as a function of the frequency $\omega$ and the distance $x$ from 
	one of the S-LL interfaces. We took $g_{\rho}=1$ (repulsive
	interactions);
	the $\delta$-functions of Eq.~(13) have been smeared, using 
	peaked Lorentzians with a width of the order the level spacing. 
	Furthermore, $Lk_F = 10^6,\:
	N_L=\rho_0/E_F,\: \chi=0$.}
\label{dosslls}
\end{figure}

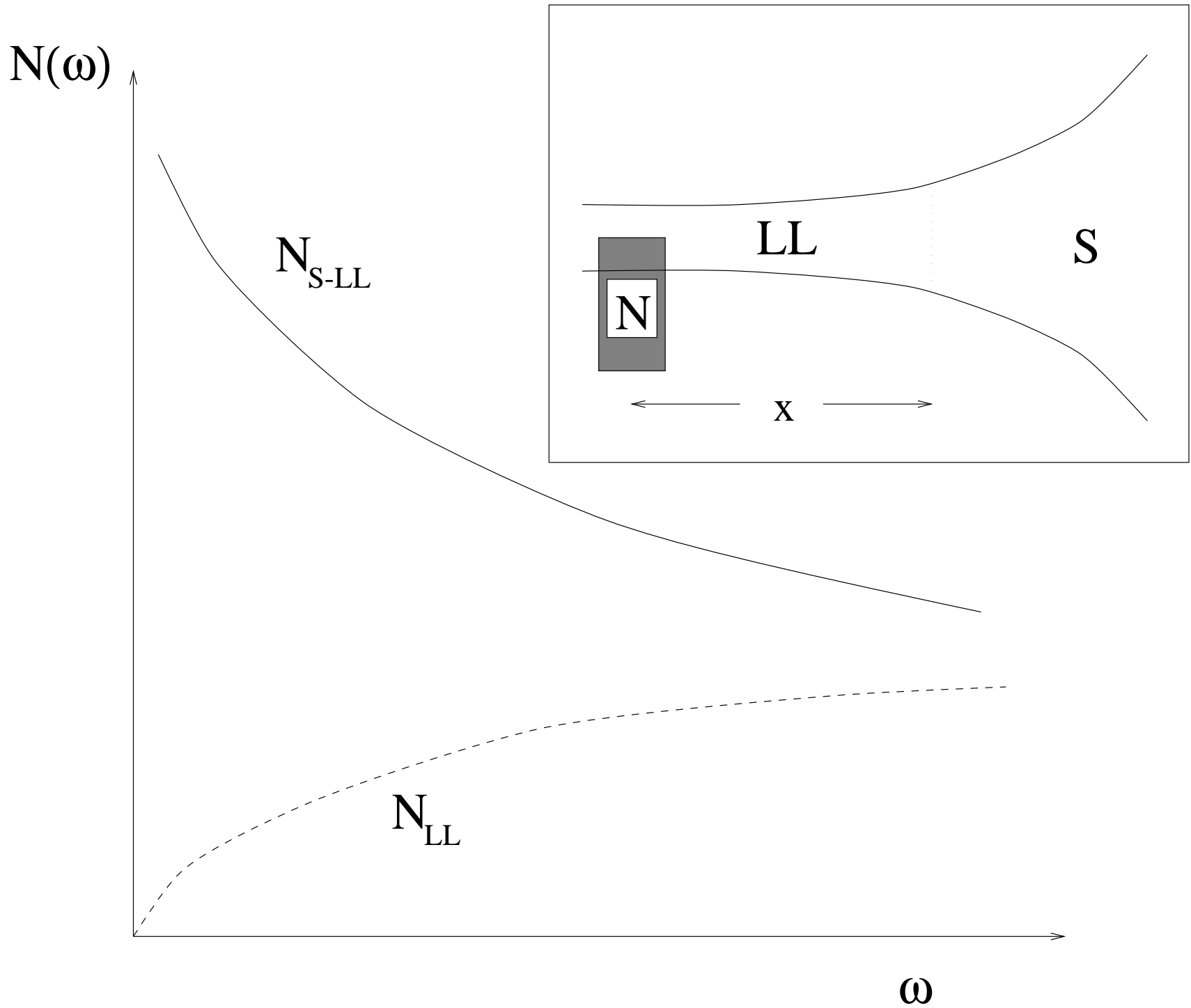
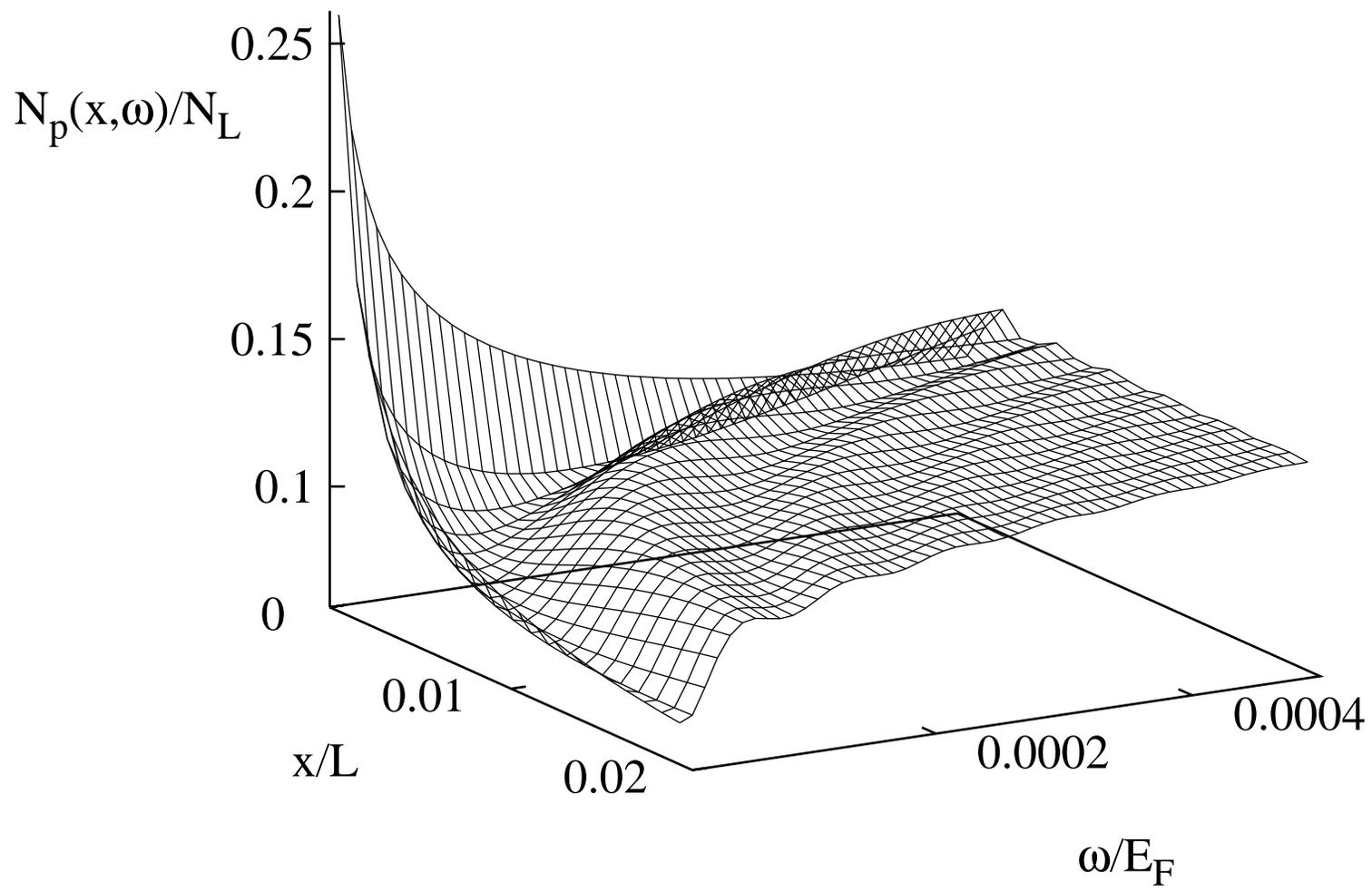
\end{document}